\documentclass[conference]{IEEEtran}
\IEEEoverridecommandlockouts

\usepackage{cite}
\usepackage{listings}
\usepackage{array}
\usepackage{balance}
\usepackage{multirow}
\usepackage{pbox}
\usepackage{url}
\usepackage{paralist}
\usepackage{xspace}
\usepackage{booktabs}
\usepackage{amsmath,amssymb,amsfonts}
\usepackage{algorithmic}
\usepackage{graphicx}
\usepackage{subfigure}
\usepackage{textcomp}
\usepackage{todonotes}
\usepackage{xcolor}
\usepackage{balance}
\usepackage{booktabs}
\usepackage{orcidlink}
\usepackage{wrapfig}

\newcommand\orcid[1]{\orcidlink{#1}~\text{#1}}
\makeatletter
\newcommand{\linebreakand}{%
  \end{@IEEEauthorhalign}
  \hfill\mbox{}\par
  \mbox{}\hfill\begin{@IEEEauthorhalign}
}
\def\tightitemize{\ifnum \@itemdepth >3 \@toodeep\else \advance\@itemdepth \@ne
\edef\@itemitem{labelitem\romannumeral\the\@itemdepth}%
\list{\csname\@itemitem\endcsname}{\setlength{\topsep}{-\parskip}\setlength{\parsep}{0in}\setlength{\itemsep}{0in}\setlength{\parskip}{0in}\def\makelabel##1{\hss\llap{##1}}}\fi}

\makeatother
\newcommand{\EndStep}{{\tt EndStep()}} 
\newcommand{\BeginStep}{{\tt BeginStep()}} 
\newcommand{\Put}{{\tt Put()}} 
\newcommand{\Open}{{\tt Open()}} 
\newcommand{\Close}{{\tt Close()}} 

\def\BibTeX{{\rm B\kern-.05em{\sc i\kern-.025em b}\kern-.08em
    T\kern-.1667em\lower.7ex\hbox{E}\kern-.125emX}}

\begin{document}
\title{Streaming Data in HPC Workflows Using ADIOS}

\author{
    \IEEEauthorblockN{Greg Eisenhauer\IEEEauthorrefmark{1}}
    \IEEEauthorblockA{\orcid{0000-0002-2070-043X}}
\and
    \IEEEauthorblockN{Norbert Podhorszki\IEEEauthorrefmark{2}}
    \IEEEauthorblockA{\orcid{0000-0001-9647-542X}}
\and
    \IEEEauthorblockN{Ana Gainaru\IEEEauthorrefmark{2}}
    \IEEEauthorblockA{\orcid{0000-0002-1375-9468}}
\and
    \IEEEauthorblockN{Scott Klasky\IEEEauthorrefmark{2}}
    \IEEEauthorblockA{\orcid{0000-0003-3559-5772}}
\linebreakand
    \IEEEauthorblockN{Philip E. Davis\IEEEauthorrefmark{3}}
    \IEEEauthorblockA{\orcid{0000-0002-2205-8268}}
\and
    \IEEEauthorblockN{Manish Parashar\IEEEauthorrefmark{3}}
    \IEEEauthorblockA{\orcid{0000-0003-0983-7408}}
\and
    \IEEEauthorblockN{Matthew Wolf\IEEEauthorrefmark{7}}
    \IEEEauthorblockA{\orcid{0000-0002-8393-4436}}
\and    

    \IEEEauthorblockN{Eric Suchtya \IEEEauthorrefmark{2}}
    \IEEEauthorblockA{\orcid{0000-0002-7047-9358}}
 \linebreakand   
    \IEEEauthorblockN{Erick Fredj\IEEEauthorrefmark{4}}
    \IEEEauthorblockA{\orcid{0000-0002-7991-4942}}
\and
    \IEEEauthorblockN{Vicente Bolea\IEEEauthorrefmark{5}}
    \IEEEauthorblockA{\orcid{ 0000-0002-5382-093X}}
\and
    \IEEEauthorblockN{Franz P\"oschel \IEEEauthorrefmark{6}}
    \IEEEauthorblockA{\orcid{0000-0001-7042-5088}}
\and
    \IEEEauthorblockN{Klaus Steiniger \IEEEauthorrefmark{6}}
    \IEEEauthorblockA{\orcid{0000-0001-8965-1149}}
\linebreakand
    \IEEEauthorblockN{Michael Bussmann \IEEEauthorrefmark{6}}
    \IEEEauthorblockA{\orcid{0000-0002-8258-3881}}
\and
    \IEEEauthorblockN{Richard Pausch \IEEEauthorrefmark{8}}
    \IEEEauthorblockA{\orcid{0000-0001-7990-9564}}
\and
    \IEEEauthorblockN{Sunita Chandrasekaran \IEEEauthorrefmark{9}}
    \IEEEauthorblockA{\orcid{0000-0002-3560-9428}}


\linebreakand\small
\IEEEauthorrefmark{1}Georgia Institute of Technology, Atlanta, GA
\\
\IEEEauthorrefmark{2}Oak Ridge National Laboratory, Oak Ridge, TN
\\
\IEEEauthorrefmark{3}University of Utah, Salt Lake City, UT
\\
\IEEEauthorrefmark{4}
Toga Networks, a Huawei Company, Tel Aviv, Israel and The Jerusalem College of Technology, Jerusalem, Israel
\\
\IEEEauthorrefmark{5}Kitware, Clifton Park, NY
\\
\IEEEauthorrefmark{6}Center for Advanced Systems Understanding, G\"orlitz, Germany
\\
\IEEEauthorrefmark{7}Samsung SAIT: San Jose, CA
\\
\IEEEauthorrefmark{8}Helmholtz-Zentrum Dresden-Rossendorf, Dresden, Germany
\\
\IEEEauthorrefmark{9}University of Delaware, Newark, DE
\normalsize
}

\maketitle

\begin{abstract}
The ``IO Wall'' problem, in which the gap between computation rate and data access rate grows continuously,
poses significant problems to scientific workflows which have
traditionally relied upon using the filesystem for intermediate
storage between workflow stages.  One way to avoid this problem in scientific workflows is to  stream data directly from producers to consumers and avoiding storage entirely. However, the manner in which this is accomplished is key to both performance and usability.  This paper presents the Sustainable Staging Transport, an approach which allows direct streaming between traditional file writers and readers with few application changes. SST is an ADIOS ``engine'', accessible via
standard ADIOS APIs, and because ADIOS allows engines to be chosen at
run-time, many existing file-oriented ADIOS workflows can utilize SST
for direct application-to-application communication without any source
code changes.  This paper describes the design of SST and presents performance results
from various applications that use SST, for feeding model training with simulation data with substantially higher bandwidth than the theoretical limits of Frontier's file system, for strong coupling of separately developed applications for multiphysics multiscale simulation, or for in situ analysis and visualization of data to complete all data processing shortly after the simulation finishes.
\end{abstract}
\section{Introduction}
As high performance computing systems have evolved to exascale and
beyond, bandwidth to the filesystem has not kept up with increases in
compute speed and memory capacity. This ``IO Wall'' problem in which the
gap between computation rate and data access rate grows continuously
poses significant problems to scientific workflows which have
traditionally relied upon using the filesystem for intermediate
storage between workflow stages.  This paper describes the Sustainable
Staging Transport (SST), a component of the ADIOS I/O system~\cite{ADIOS2} that
avoids the IO wall by streaming data directly from consumers to
producers in HPC workflows.  SST is an ADIOS ``engine'', accessible via
standard ADIOS APIs, and because ADIOS allows engines to be chosen at
run-time, many existing file-oriented ADIOS workflows can utilize SST
for direct application-to-application communication without any source
code changes.

SST builds upon prior research efforts, including Flexpath\cite{flexpath}, a component of an earlier version of ADIOS which coupled analytics to applications such as LAMMPS and GTS and demonstrated on the Titan supercomputer faster end-to-end completion times than processing data through the file system. 
Flexpath differed from prior work, such as Dataspaces\cite{dataspaces} in that it allowed direct communication between data producers and consumers in order to avoid the extra data movements involved with publishing data to an external broker.  
While Flexpath was a useful proof of concept, it lacked critical features, such as the ability to use RDMA networks for data movement, dynamic connection and disconnection for data consumers and robustness to failures.   
In contrast, SST was designed to be a practical and stable communications layer allowing direct coupling between applications running at the highest scales on HPC resources.

\section{Goals and Objectives}
\begin{figure*}[tb]
    \centering
    \includegraphics[width=1.2\columnwidth]{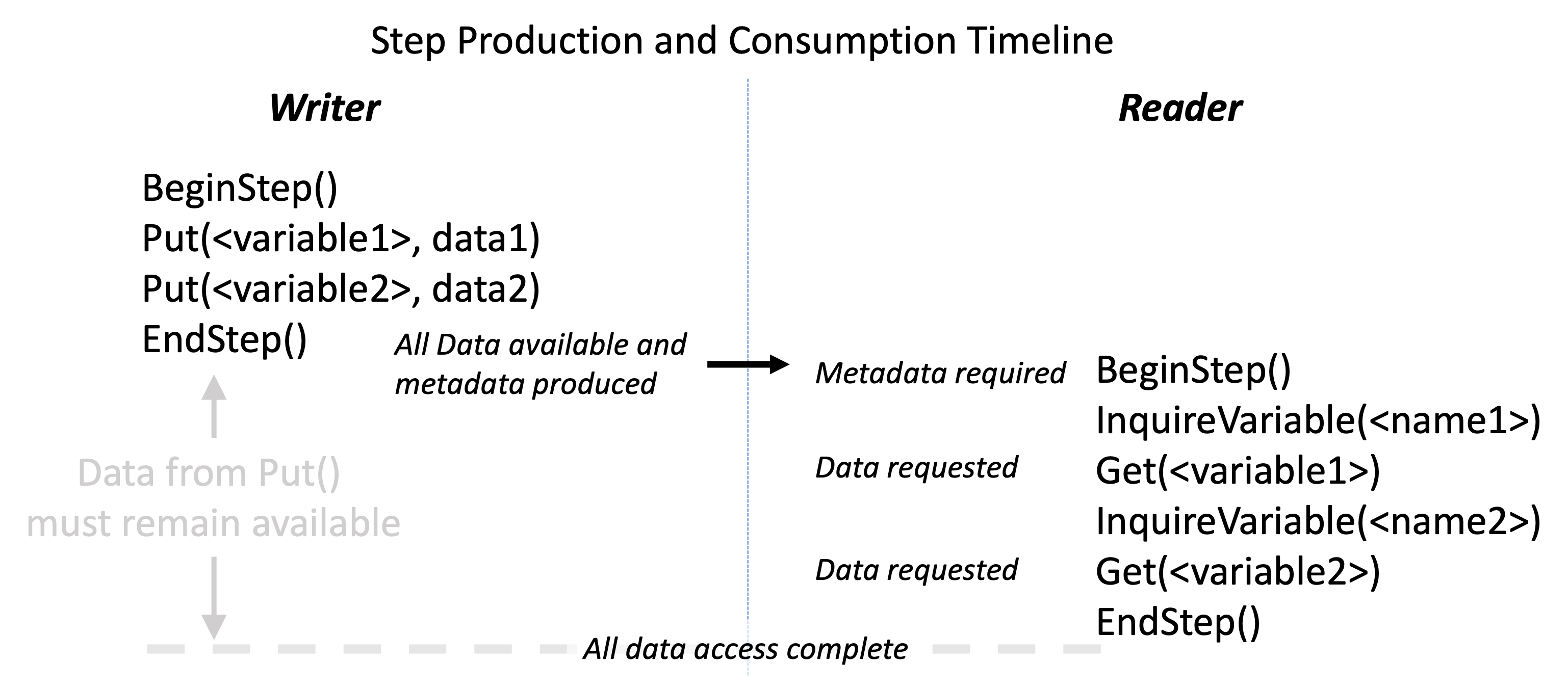}
    \caption{Reader- and writer-side timelines for a single step. This figure shows only a single reader and writer rank, but in fact each side may consist of thousands of MPI ranks.}
    \label{fig:timeline}
\end{figure*}

Before diving into the details of SST, we must introduce the environment in which it operates.
ADIOS is a timestep-oriented HPC I/O framework which creates
self-describing output files and supports complex data structures such
as N-dimensional global arrays decomposed in arbitrary ways across
writer and reader ranks.  In this context "timestep-oriented" means
that on the writer side, I/O is divided into distinct steps and in
each step each writer rank may write all or a portion of an ADIOS
"variable".  ADIOS readers then can read the written file
step-by-step, and each reader rank may request all or a
portion of each written variable.  
ADIOS does not assume any kind of fixed or even regular layout pattern for data produced by the writer. While the number of dimensions that a global array has cannot be changed after it is defined, the size of the overall global array can change from step to step, and the portion of the array written by any individual writer rank can also change.  
The ADIOS reader side is similarly flexible in that each reader rank can query the characteristics of the variables written on each individual step and decide on a step-by-step basis exactly which portions of the written data that it will read.

While the ADIOS API was developed for efficient large-scale HPC IO to
files, its step-oriented nature makes it adaptable to support
streaming data between running HPC applications with the simple
expedient that data producers use the writer-side ADIOS API and data
consumers use the reader-side API while the SST engine arbitrates the
data exchange to preserve file-like semantics for both sides.

The goal of SST was to fully support normal ADIOS writer-side and reader-side semantics as well as desirable streaming features. These goals include:
\renewcommand{\labelenumi}{Goal \arabic{enumi}:}
\begin{enumerate}
\item \label{itm:parallel} writer and readers are parallel (multi-core) programs
\item \label{itm:change} writer data geometry and reader data selections can change on each step
\item \label{itm:impede} ensuring that readers don't impede writer progress or cause its failure
\item \label{itm:dynamic} dynamic connection and disconnection of reader applications
\item \label{itm:multiple} support for multiple simultaneous reader applications
\end{enumerate}
\renewcommand{\labelenumi}{\arabic{enumi}}


The intersection of these goals and ADIOS semantics have some important implications for an ADIOS streaming data system.  
With respect to Goal~\ref{itm:change} above, that the full details of what is being written aren't available until the writer-side ADIOS \EndStep call, yet the reader must acquire that information during its \BeginStep call is particularly impactful.  
This timing relationship results in a situation as depicted in Figure~\ref{fig:timeline}.  
There are several things that are implicit in this diagram, one being the separation of {\it data} (the contents of the application data block supplied to \Put and {\it metadata} (information like variable size, block count and other ADIOS-level information that is required for ADIOS semantics). 
ADIOS metadata handling in file environments is described in prior work\cite{RanjanDAOS} and there are similarities in an ADIOS streaming environment, particularly with respect to all reader ranks requiring access to the metadata from every writer rank.  
While this characteristic of ADIOS I/O has several consequences, the implications to the timing relationship between production and consumption of a step are the most important for streaming data.  
In particular, ADIOS semantics require that data provided by the writing application in \Put calls can be overwritten or destroyed as soon as 
\EndStep returns.  
However, because the full metadata for the step must be available to the reader when reader-side \BeginStep returns, any reader-side decision-making and requests for data can only occur after that.  To summarize: 
\begin{figure*}[tb]
    \centering
    \includegraphics[width=\columnwidth]{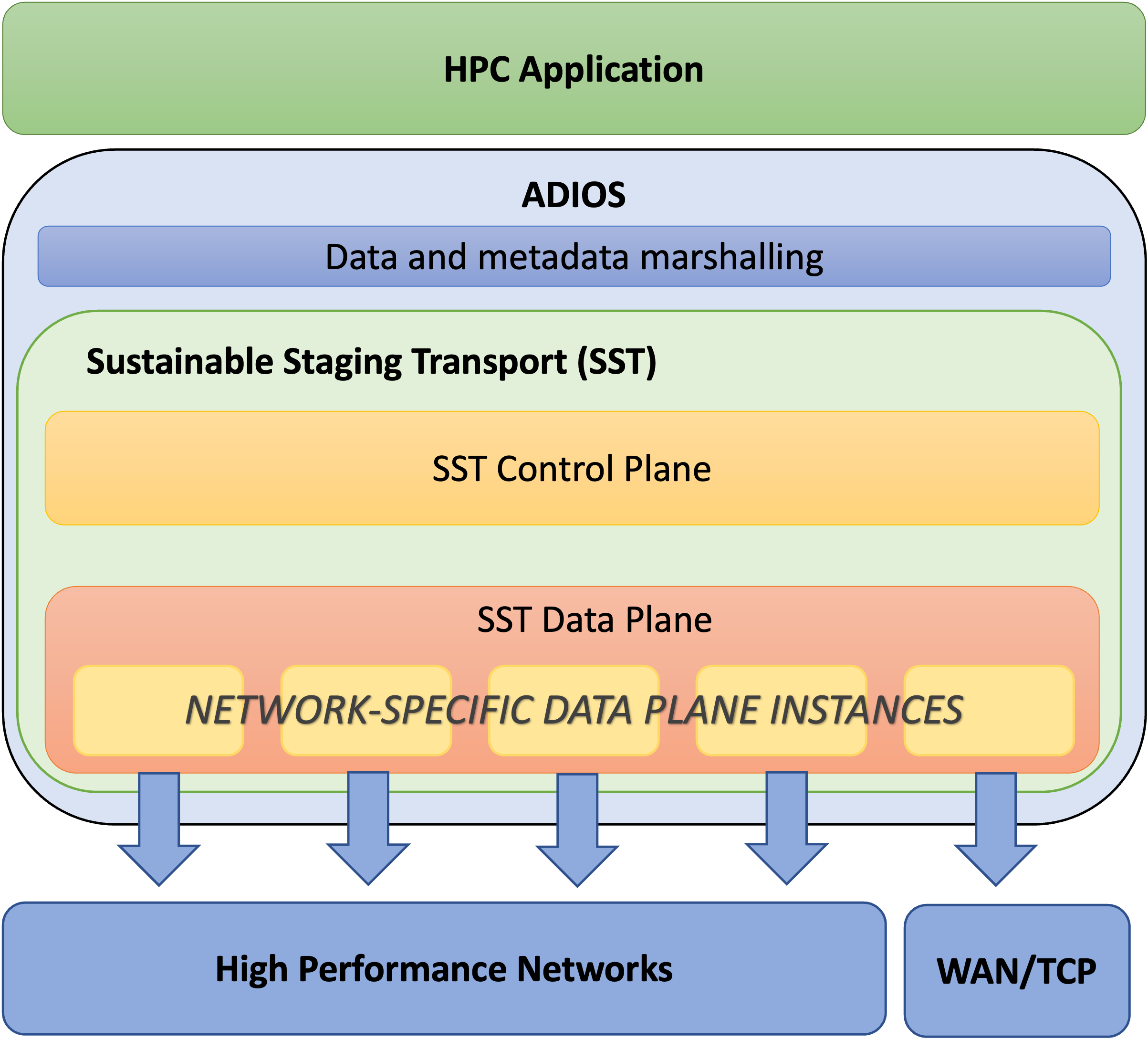}
    \caption{Overview of SST in its role as an ADIOS engine, including internal architecture.}
    \label{fig:SST_architecture}
\end{figure*}
\begin{tightitemize}
    \item metadata can't be created until the start of writer-side \EndStep
    \item application data buffers are released by the return of writer-side \EndStep
    \item reader requests for data can't proceed until metadata is received.
\end{tightitemize}
This situation constrains implementation options for ADIOS streaming.   Writer applications in ADIOS are free to destroy the application data buffers supplied in \Put as soon \EndStep returns, so if we want to avoid copying that data we have to either {\bf a)} stall the return of writer-side \EndStep until the end of reader \EndStep so that data requests can succeed, or {\bf b)} somehow transmit all writer data to those readers who will need it so that \EndStep can return.  Here, option {\bf a} is essentially making the reader to be fully synchronous with the writer and essentially conflicts with item~\ref{itm:impede} in our list goals above (by letting readers impede writer progress).  
On the other hand, option {\bf b} isn't really possible because it requires foreknowledge of reader requests which isn't available in ADIOS semantics.  
There are ways around that, such as sending all the data to a intermediary for later distribution (the approach used by broker-based solutions), or possibly simply sending all the data to every reader.  
The former we rejected as being duplicative of prior approaches like DataSpaces. The latter, sending all the data to every reader, is obviously impractical at scale.

With ADIOS semantics rendering design options that avoid data copies  impractical or undesirable, SST focused on a queue-oriented model of data streaming in which application output data is copied in \EndStep and queued on each writer node until it can be consumed by reading applications.  
While data copies are often undesirable in HPC I/O, buffering data in SST has opened up a wide variety of opportunities, including allowing background I/O to continue while the writer application continues to compute, enabling scheduling I/O to use the network in application compute phases, 
and the introduction of special controls to manage overall writer-side memory consumption.  
It also allows SST to support dynamic connection and departure of read clients and to support multiple readers, as specified in Goals~\ref{itm:dynamic} and \ref{itm:multiple} above. Note, that many applications at large scale have data in memory organized in a different way than what is published as output, and ADIOS has interfaces (e.g. Span buffer allocation, and GPU-aware I/O) to allow the user to directly prepare the output data in ADIOS buffers. This was true for the particle in cell codes described in the application section, and is true for code coupling projects where the developers add new data products specifically for the purpose of exchanging state between different applications. 
Given this baseline queue-oriented model, the next section details the overall architecture of SST and how it achieves data streaming with high levels of performance.

\section{Architecture and Implementation}

\begin{figure*}[tb]
    \centering
    \includegraphics[width=\columnwidth]{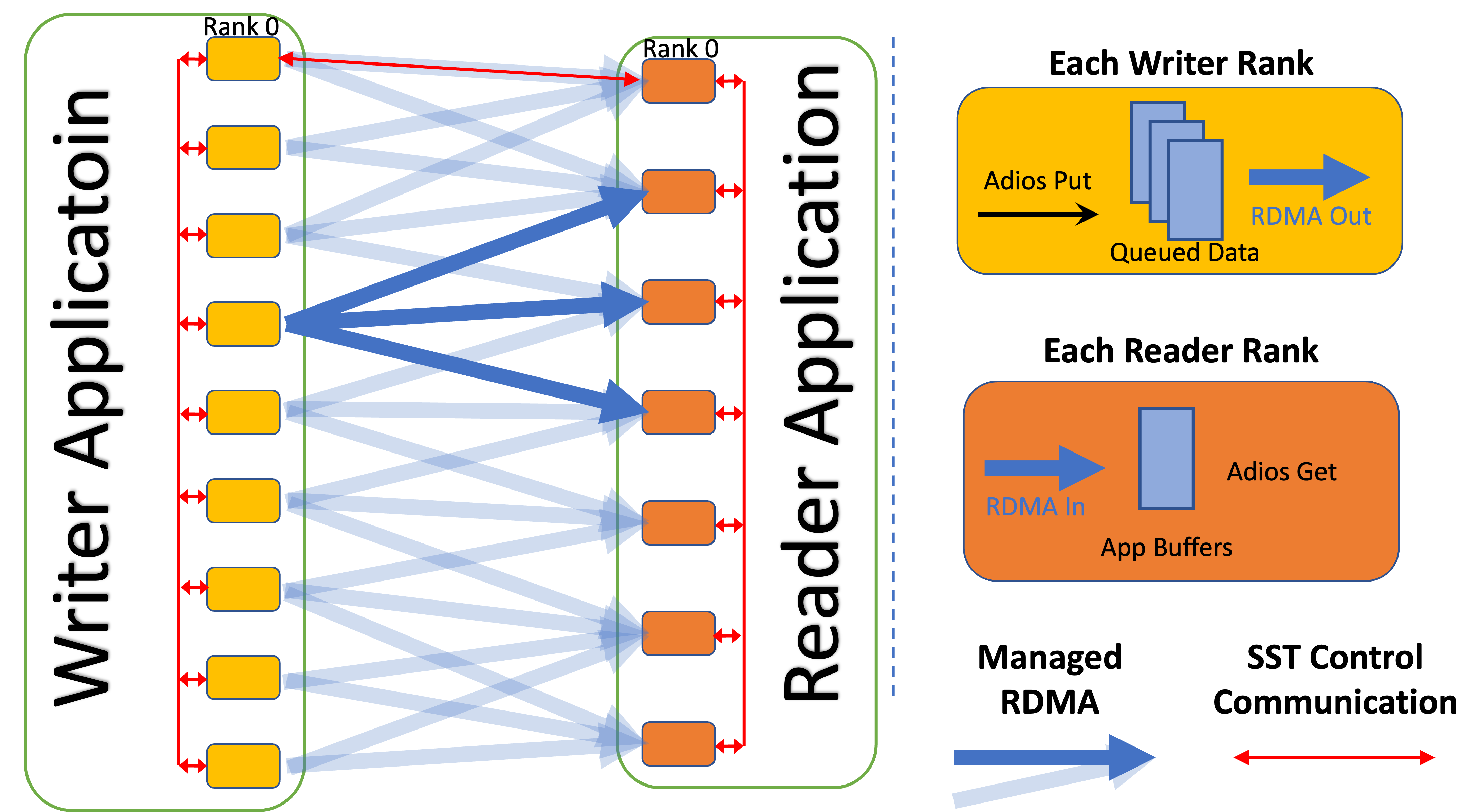}
    \caption{Depiction of communicating SST applications.}
    \label{fig:SST_cohorts}
\end{figure*}
The overall architecture of SST is shown in Figure~\ref{fig:SST_architecture}.  As an ADIOS engine, it is internal to ADIOS and interfaces with the application through the normal ADIOS APIs.  It also leverages the data and metadata marshaling methods developed for the ADIOS file engines.  In particular, SST originally utilized variations of the BP3 file engine marshalling methods.  However as detailed in~\cite{ecp_paper}, the BP5 serializer was developed as a more efficient and flexible serialization method that could be directly shared between file and streaming engines like SST.  
While the internal engineering of ADIOS places the serializer inside the Engine, the architecture and operation of SST are largely independent of the data and metadata serialization (or marshalling), so Figure~\ref{fig:SST_architecture} shows marshalling outside of SST and discussion will follow that convention as much as possible.   On the bottom of Figure~\ref{fig:SST_architecture} we see interfaces to high performance RDMA networks.  The ability to use the high performance networks on HPC clusters is key to SST performance on the HPC clusters of interest, though we do include a TCP/IP-based interface both for development purposes on non-RDMA systems and for use in wide-area-networking situations.  

The green Sustainable Staging Transport box contains a high-level {\em Control Plane} which handles things like metadata delivery and step advancement, and a lower level {\em Data Plane} devoted to high speed data delivery.  For clarity, we note that SST does not rely upon MPI for inter-job communication.  This work is aimed at streaming data between large-scale MPI codes that may have been previously developed to run separately and communicate via files and usually not ready to run in MPI's MPMD mode under one world communicator.  
MPI inter-communicators are also not always available or functional, so SST's design revolved around direct access to RDMA networks for data transfer, with MPI-based inter-application communication as one of several fallbacks.

\subsection{Overall Operation}
While more detail follows below, it's useful to first present an overview in order to place later discussions in context.
Given the architecture of Figure~\ref{fig:SST_architecture}, Figure~\ref{fig:SST_cohorts} provides a depiction of data flows during a step. 
On the left side, SST communicates control and metadata between the writing ranks (in writer \EndStep) while application data (handed over to ADIOS in \Put) remains queued on its originating rank.  Metadata and additional information is aggregated to writer Rank~0, where it is communicated to reader Rank~0 who then shares it with the other reader ranks.  As discussed above and in \cite{RanjanDAOS}, ADIOS metadata consists of detailed information about the ADIOS Variables written in a step, including the overall geometry of the variable and the geometry of the each written data block.  This information is required for Reader-side \BeginStep, after which application code can issue queries to that information and make read requests.  Many Data Planes also require writer-generated Data Plane-level information to be made available to the reader side to enable these transfers, so information is gathered and distributed by the Control Plane along with the ADIOS-level metadata.  Here reads are represented as right-moving arrows which may be RDMA pull operations transferring data from writer buffers into reader destinations.  

\subsection{Data Planes\label{sec:data_plane}}
Of the two internal components to SST identifed in
Figure~\ref{fig:SST_architecture}, the Data Plane is the simpler, so
we'll describe it first.  The Data Plane's contract with the higher
layers of SST is relatively basic.  On the writer side, buffers
containing data for Step {\tt N} will be ``registered'' with the data plane instance on the rank on which they are produced (and as noted above, the data remains queued on the writer until no longer needed).   
Later, a reader-side data plane instance will get a ``read request'' consisting of a Step number, writer rank, starting byte offset and length of data to be read and a destination buffer in which it is to be placed and the Data Plane is expected to deliver the data.
In an HPC environment its to be expected that the data movement is one of the most critical aspects of performance and must be accomplished on the highest-performing networks available, but SST leaves those details to custom dataplanes designed to exploit the hardware.  
In terms of support from upper layers, in addition to the writer-side data registration mentioned above, data planes are also notified when registered data can be released (I.E. when all readers have moved past that Step and no further read requests will be incoming) so that resources can be released.  The data plane registration operation can also return network-specific access information (like MR Keys for pinned memory regions used in RDMA communication) about the queued data.  This information is not interpreted by the control plane
but is gathered and provided to the reader-side data plane for use in servicing read requests.

At the beginning of each reader-side Step, the reader side Data Plane is provided the per-writer information gathered from the writer Data Plane in the registration operation.  
Once initialized with this information, the Data Plane will receive data read requests passed down from upper layers as described above.  
Read operations are split into an asynchronous request, which returns a handle, and a wait operation which must block until the data has been delivered to the destination.
This split allows upper levels to issue all the reads that are required while still allowing the dataplane to pipeline and/or parallelize the operations in whatever way maximizes throughput for the actual network.

\begin{figure*}[htb]
    \centering
    \includegraphics[width=.8\columnwidth]{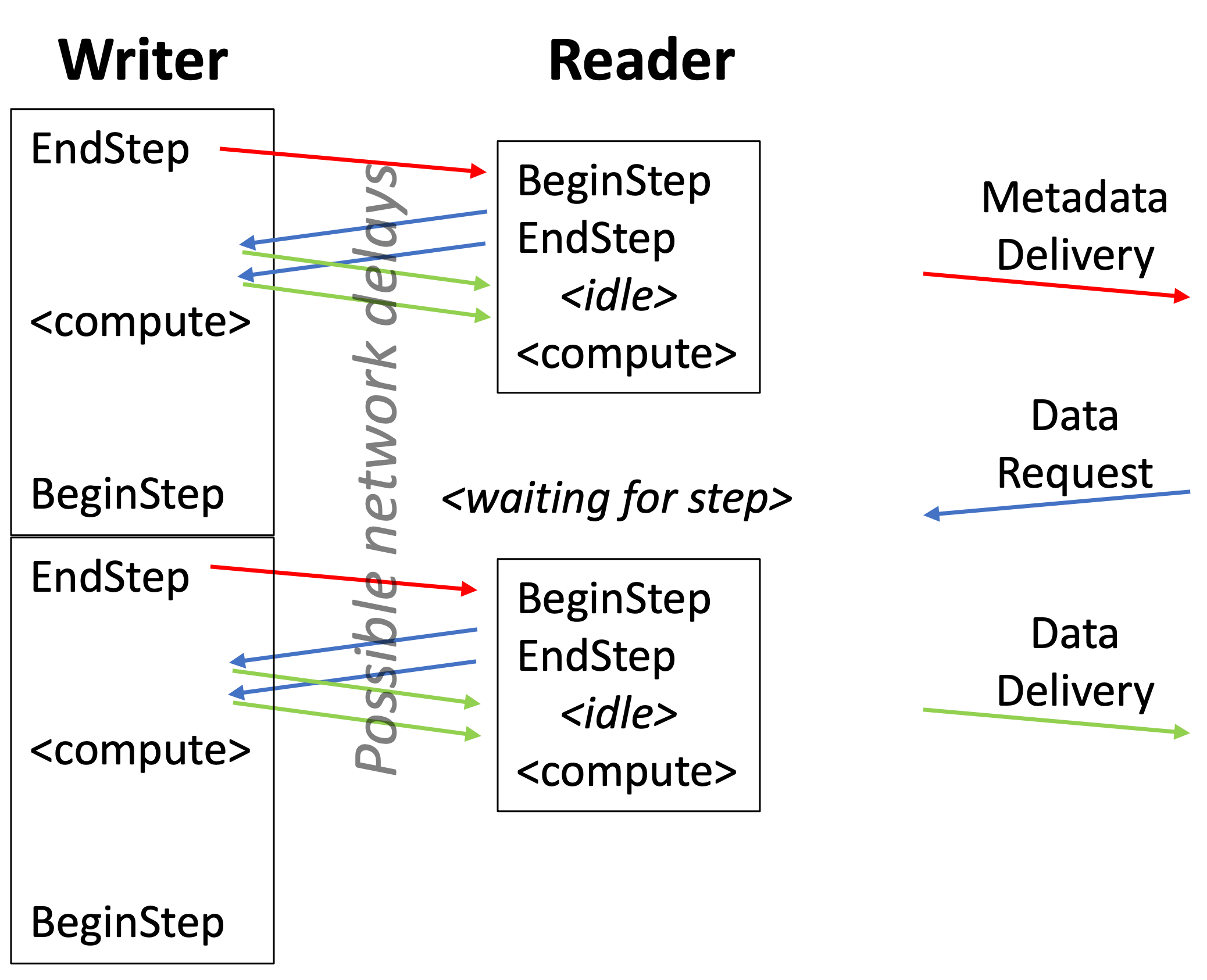}
    \caption{Request / response timeline in Dataplane communications. All communication take place in the writer concurrently with the writer running its compute task. }
    \label{fig:datatimeline}
\end{figure*}

As depicted in Figure~\ref{fig:SST_architecture}, SST has a number of network-specific data planes it can use in different situations.  
In general, the writer picks the highest-performing data plane available to it, but can configured to use a specific data plane with the SST {\bf DataTransport} parameter.  Currently extant data planes are described below.
\subsubsection{EVPath IP-based Data Plane}
SST's low-end dataplane, called EVPath because it's based on the EVPath library\cite{EVPath} is tcp-based and intended primarily for debugging and
application development. Its per-reader contact information includes
an IP and port number upon which the writer data plane listens for
requests.  As shown in Figure~\ref{fig:datatimeline}, a read request
made on the reader side results in a {\tt ReadRequest} message being
sent from that particular reader rank to the writer rank where the
data resides.  That message is handled by a background network-handler
thread on the writer, which then sends a {\tt RequestResponse} message back to the requesting reader.  From the reader's point of view, multiple requests may be outstanding simultaneously, and a read wait operation is simply waiting on a particular response message.  
The TCP data plane is adequate for debugging and may serve in WAN environment where no high performance network is available, but its reliance on individual reader-rank-to-writer-rank connections means that its scalability is limited by the number of sockets/file descriptors available on nodes. 
The EVPath data plane can also be configured to use a Reliable UDP protocol which doesn't suffer from the same socket limitations, but neither data plane can match the performance of Data Planes based on RDMA-networks.

\subsubsection{LibFabric Data Plane}
Our initial high-performance data plane was based on the LibFabric communication framework, an implementation of the Open Fabric Interface (OFI)\cite{grun2015brief}. OFI abstracts operations such as messaging and bulk transfer over high-performance network fabrics using RDMA. 
In its default mode of operation, RDMA-accessible buffers are created as part of the timestep registration process. All buffer information on the writer side is contributed to the timestep metadata, and these metadata are provided to the reader to use during reads. Each read is mapped into one or more RDMA get operations which complete asynchronously, only blocking the reader while any outstanding requests complete during \EndStep. 
In this mode, the writer is passive, performing no RDMA operations other than bootstrapping the network state and registering timesteps, unless other writer side activities in the libfabric layer are required to make progress on all outstanding data transfers initiated by the readers. 

\subsubsection{UCX Data Plane}
Because LibFabric isn't functional across all HPC networks, we also have a data plane targeted to the Unified Communication X (UCX) framework.  In many ways, the UCX data plane functions similarly to the LibFabric data plane, just with a different higher-level RDMA wrapper library.  Both provide the ability to map memory and perform RDMA get operations.  Between UCX and LibFabric together, they cover most HPC vendors' networks.

\subsubsection{MPI Data Plane}
In situations where the LibFabric or UCX data planes might not function or be performant, SST does have an MPI-based data plane that uses MPI {\em inter-communicators}.  In particular, this data plane uses control plane services to send {\tt ReadRequest} and {\tt RequestResponse} messages like those described in the TCP-based dataplane, except that the {\tt RequestResponse} {\em does not carry the actual data.}  
Instead, receipt of a ReadRequest message on the writer triggers an {\tt MPI\_Send()} operation and the {\tt RequestResponse} arrival results in an {\tt MPI\_Recv()}, so that the data travels via MPI and only the control happens via TCP.  Because MPI is generally highly optimized on HPC platforms, this approach can be highly effective.  However, it comes with some restrictions:
\begin{itemize}
    \item The version of MPI in use must support inter-communicators,
    \item Because data transfer operations happen in the network handler thread, the application {\em must} initialize MPI with threading, using {\tt MPI\_Init\_thread()} and specifying {\tt MPI\_THREAD\_MULTIPLE} as the level of threading support required.
\end{itemize}
As a result of using a feature of MPI that is not widely used and tested, currently only MPICH-based versions of MPI work properly for this data plane. As shown later in the application section, this solution is scaling and performing very well up to the full scale of Frontier. 

In some cases, using thread-safe MPI incurs more overhead than non-thread-safe, so the application may suffer additional overhead in MPI operations outside ADIOS but we haven't found an ADIOS application yet that suffers from this. 

\subsection{Control Plane\label{sec:control_plane}}
The purpose of the control plane is to orchestrate the overall scheme depicted by Figure~\ref{fig:SST_cohorts}, managing the initial connection of readers to writers, controlling the advancement of Steps, delivering the Step information generated by the writer to the reader, and other sundries such as error and shutdown handling.
The Control Plane relies upon MPI collective operations within the ranks of a reader or writer, and a third-party library, EVPath\cite{EVPath}, for communication between reader and writer cohorts.  
Generally the operations that are collective in SST are those that are also collective in the ADIOS file engines, such as \Open, \Close, \BeginStep and \EndStep and it is within those calls that SST functionality that requires the cooperation of all ranks is implemented.  For communication between reader and writer cohorts, the EVPath libraries essentially provides simple message passing and background network handling.  By default, EVPath uses TCP/IP sockets and for control plane purposes, SST only requires a connection between the rank 0 of any writer and reader.  However, the control plane also offers supporting message passing services to data planes, which may require as much as all-to-all connectivity between the reader and writer ranks, depending upon the data request pattern.  As noted in Section~\ref{sec:data_plane} above, this may push resource limit WRT per-process file descriptors, so EVPath also has a reliable-UDP transport which limits resource consumption at the cost of doing user-level reliability.

On the writer-side, the action most pertinent to SST starts in \EndStep, at which point the upper ADIOS layers {\em on each rank} separately provide SST with a local data block and local metadata block.\footnote{This is a simplification for presentation.  There are also attribute blocks, and for BP5 meta-metadata blocks but their presence and handling doesn't change the overall operation.}  EndStep() handling involves the following basic sequence of events:
\begin{enumerate}
\item each SST rank receives a local data block and local metadata
  block via SST's {\tt  ProvideTimestep()} interface
\item each rank registers the local data block with the data plane and receives back registration information
\item all ranks participate in an MPI broadcast from rank 0 of SST status information (detailed later)
\item local metadata and dataplane registration information is gathered to rank 0 via an MPI collective to form the full metadata
\item rank 0 sends a {\tt ProvideMetadata} message containing the full metadata to rank 0 of the reader
\item \EndStep returns but the local data and metadata remain queued in memory on each writer rank
\end{enumerate}

On the reader-side, the {\tt ProvideMetadata} message will arrive asynchronously and be queued, but when the reader is blocked waiting for the next timestep, this message arrival will wake the main thread of rank 0 to continue processing.  
In particular reader-side BeginStep(), rank 0 will check for queued
metadata for the next step and if it has not yet arrived it blocks
pending arrival of a {\tt ProvideMetadata} message.  
Once it has metadata, rank 0 uses an MPI collective to broadcast it to the other reader ranks.  
Non-zero ranks enter the MPI broadcast collective immediately on BeginStep() to receive metadata, so they implicitly block along with rank 0 if metadata has not yet arrived.  
Once the broadcast has completed, each rank "installs" the metadata, using it to create the available ADIOS variables and populate them with available block information, after which BeginStep() returns.

Between Begin/EndStep pairs, applications may do ADIOS Get() operations, each of which results in one or more data plane read requests.  The use of a synchronous Get(), or a call to PerformGets() (both non-collective calls) results in calls to the local data plane to wait for all pending dataplane requests to complete.  
The control plane doesn't directly participate in these operation, but passes requests and responses between the upper layers and the data plane below.
EndStep() also results in a wait for completion of all pending read
requests on the local node, but afterwards each reader rank
participates in an MPI Barrier().  When all reader ranks have passed
this barrier, rank 0 sends a {\tt ReleaseStep} message to writer rank
0, EndStep() returns and the reader-side Begin/EndStep sequence can
begin again.  The {\tt ReleaseStep} message sent to writer rank 0 is received asynchronously and queued.  
When the writer next enters \EndStep, the information that Step {\tt N} can be released is contained in the ``SST status information'' mentioned in the writer side EndStep sequence above, which enables each writer rank to inform its data plane that we're done with that step's data and resources related to it can be released.  Final release of the queued data and metadata completes the step cycle on the writer side.

\subsubsection*{Connections and Multiple Readers}
The prior section describes the actions of each side of the control plane on a step, but first we have to establish a connection between applications.  As per Goal~\ref{itm:dynamic} above, SST should also support dynamic connection and disconnection.  In practice, there are several parts to this process.  First, any prospective readers need contact information for writer rank 0.  Since the control plane is a TCP- or RUDP-based transport, this contact information is mostly an IP and port number, but there is also information identifying the individual stream on the contact process.  By default, SST writer's place this information in a small file in the filesystem where it can typically be found by readers on the same cluster.  In situations like WAN where a shared filesystem is unlikely, SST provides a "screen" registration mechanism where the writer outputs contact information as a base64-encoded string.  In this mechanism, readers prompt for input of contact information, providing an opportunity for the string to be cut and pasted between windows.  More cloud-based contact mechanisms are also possible but not yet implemented.

Regardless of how the contact information is acquired, the first part
of the connection process involves all reader ranks blocking in \Open
while reader rank 0 sends a message to writer rank 0 and waits for a
response.  When the message arrives at writer rank 0 it is read by the
network handler thread and processed in the next writer collective
operation, which might be writer \Open if the writer is waiting for readers before continuing or \EndStep if the writer is mid-run.  In either case, the writer and reader engage in several handshake steps where they exchange per-rank control plane and data plane contact information, so that every reader rank and ever writer rank can contact each other if necessary.  

As per above, multiple readers can be paired simultaneously to a single writer, each progressing at its own pace.  
In this case, reference counts are associated with the data and metadata blocks so that none are released before all readers have consumed the data.  Writers and readers are differentiated in what happens to their peer upon close or failure.  Readers can exit the stream at any time by simply calling Close() and SST will clean up writer-side resources allocated to that reader allowing the writer to proceed with whatever readers remain (if any).   Writer Close() will block until all readers have consumed and released all remaining queued Steps.  Writer failure will cause all readers to see an EndOfStream indication on their next BeginStep(), or potentially see an exception raised in Get() or EndStep() if remote read operations fail because of writer unavailability.

\subsection{Extended Functionality\label{sec:extended}}
There are some applications for which the basic SST functionality described above is not a natural fit, or need to be modified to address specific concerns.  This section details a few modifications that have been implemented to extend that functionality.
\subsubsection{Queue Management\label{sec:queue_management}}
While ADIOS' queue-based Step management approach frees the writer from lockstep synchronization with readers, it does raise concerns about the number of steps that might end up queued on the writer and the resulting memory demands.  
The obvious solution here is the ability for the writer to set a limit on queue size, accomplished with the ADIOS engine parameter {\bf QueueLimit}, however, what happens when the writer reaches the queue limit depends upon application requirements.  
SST's default queue full policy is to block the writer until readers release a timestep, freeing up room in the queue.  
This conflicts with our stated Goal~\ref{itm:impede} above, but is necessary if the application can't lose data.   
Alternatively, the queue full policy can be set to discard steps so that the writer can continue operation without delay.  In either case, the queue length is examined in the \EndStep collective so that all writer ranks behave synchronously.  
In the case of a discard decision, the step that would have been added to the queue by \EndStep is the one that is discarded.  The somewhat more intuitive choice of discarding the step at the {\em front} of the step queue (the oldest), was rejected because it would require distributed consensus between the writer and all readers, possibly slowing overall progress.

In general, Steps are released when readers have consumed them, but if a writer is running with no readers attached Steps might be released immediately upon production.  This is problematic for some SST use cases, such as temporarily connecting a visualization tool to a long-running simulation to monitor progress.  
In that circumstance the application would really like to have a queue of the most-recently produced data waiting for it upon connection, rather than having to wait for new steps to be produced.  To support this, SST also has a "reserve queue" with a {\bf ReserveQueueLimit}.  
When this limit is non-zero, steps that have been released from the normal Step queue are moved to the reserve queue and retained specifically for the purpose of being available to be served to newly-connecting read clients.  Unlike the policy for the regular Step queue, when the reserve queue limit is reached, the oldest step (head of the queue) are discarded as new ones are added.

In addition to writer-side step management, SST also offers the reader
flexibility beyond simply consuming each step sequentially and
blocking in \BeginStep waiting for the next.  As noted above, Steps
"arrive" at a reader when a {\tt ProvideTimestep} message is received and queued by reader rank 0.
An optional {\em timeout} parameter to \BeginStep that allows reading applications to proceed with other work if a Step isn't available either immediately or within the specified time interval.   
Additionally a reader-side engine parameter {\bf
  AlwaysProvideLatestTimestep} tells SST that, in the case where the
metadata for more than one Step may be available, this reader always
wants the most recent.  If this parameter is specified to SST, the
arrival of a second {\tt ProvideTimestep} message when another is
already queued at the reader results in the new message replacing the
queued one, essentially discarding the older Step without processing.
Here, the SST reader will immediately send a {\tt ReleaseStep} message for the discarded Step to the writer so that writer-side resources can be released.

\subsubsection{Preloading in the Dataplane\label{sec:preload}}

As noted earlier in this paper, the flexibility that ADIOS allows its application require a data queueing-based model in which data doesn't move until it is requested by specific readers, resulting in a request/response pattern at the Dataplane level.  
In circumstances like those depicted in Figure~\ref{fig:datatimeline},
where the workflow's bottleneck is the writing side, any delays
potentially introduced by the request/response pattern (like the {\tt <idle>} time depicted in the reader) are immaterial to workflow completion time.  
However, if the workflow is network-bound or constrained by reader-side computation, these request-response delays may be a performance obstacle.
While writer-side bottlenecks where this is not an issue are likely more common in HPC environments, ADIOS has a feature to mitigate this in circumstances where it might be a problem.  
In particular, since this situation arises from ADIOS' flexibility in
allowing readers and writers to change pattern of their data access on
every step, ADIOS has a mechanism for applications to "lock" those
patterns, declaring that once established they will not change.  On
the writer side, this is accomplished by calling {\tt LockWriterDefinitions()} 
and on the reader side with {\tt LockReaderSelections()}.  
If both the reader and writer call these the lower level data flow pattern will not change.  This information is communicated to SST which allows some Dataplanes to enter {\bf LearnedPreloadMode}, in which data is immediately sent from writers to the readers which are known to require it without waiting for a request.  
Once a data plane is notified that the data production and consumption geometries are fixed, it can figure out where data must be delivered by watching the read requests in the next timestep. Thereafter it can deliver that data with the same distribution pattern into a queue on the reader side without waiting for a request.

Currently two of our data planes implement {\bf LearnedPreloadMode}, the EVPath and LibFabric data planes.  In the case of the EVPath data plane, the writer ranks track which reader ranks they receive read requests from and once that pattern is known, {\tt PreloadData} messages are sent from each writer to predicted readers carrying the actual data.  This send happens in writer-side \EndStep.  When these messages arrive at the reader, they are handled asynchronously and queued by the background network thread. When the reader-side DP receives a ReadRequest() these queued messages are searched so that the ReadRequest can be satisfied with a local memcpy, avoiding the request/response pattern mentioned above.  However, this approach has the downside that it consumes memory on the reader-side roughly equivalent to the amount of data read by each reader multiplied by the value of the {\bf QueueLimit} parameter.

\begin{figure}[htp]
    \centering
    \includegraphics[width=\columnwidth]{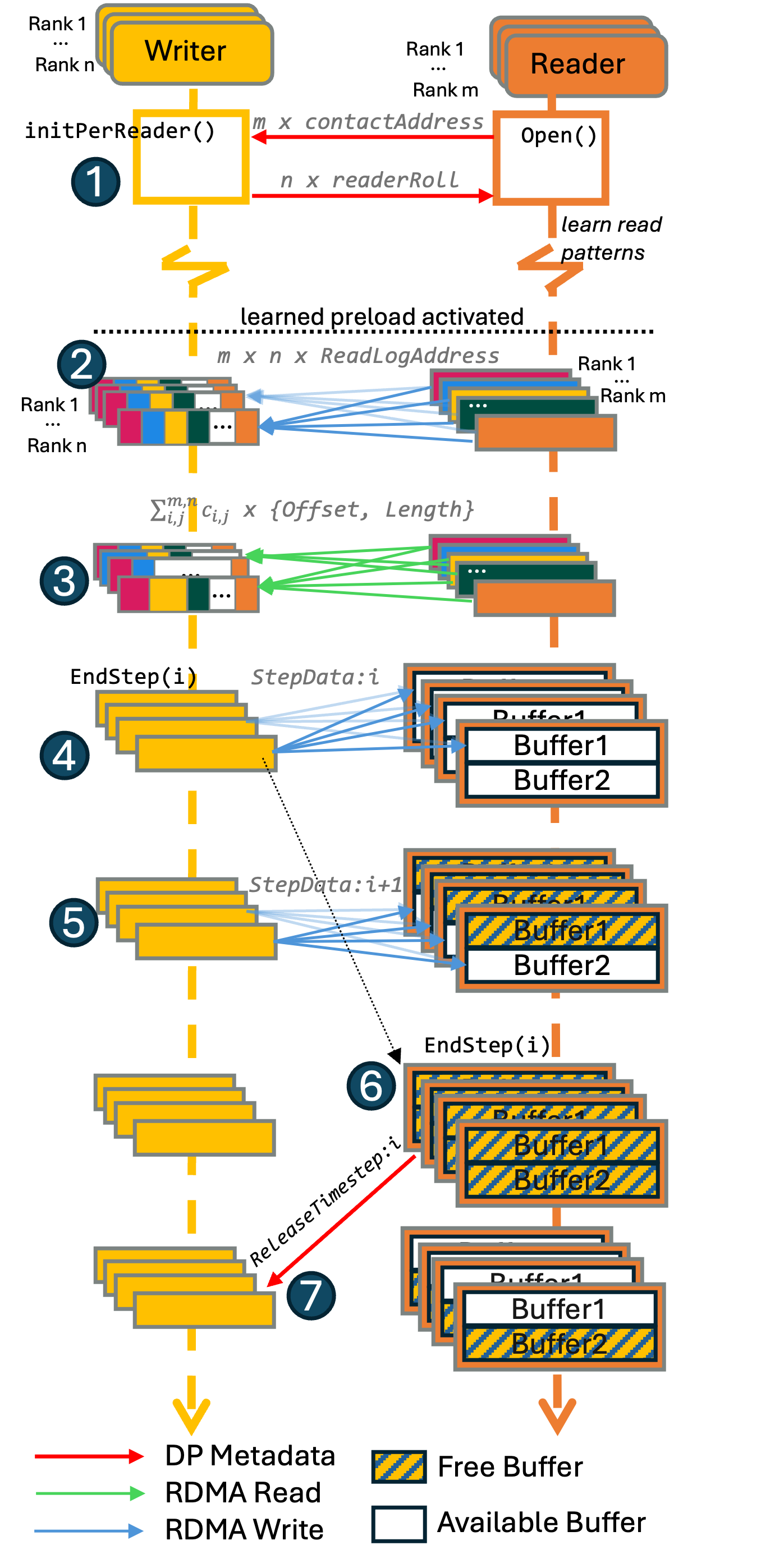}
    \caption{LibFabric data plane operation: at init, 1) each writer rank advertises a writable buffer for future read pattern data to be included in collective metadata. 2) Once preload is activated, reader ranks advertise readable buffers that contain access patterns to the writers ranks from which they have read, which then 3) ingest the access patterns; 4) the writers then push the next timestep as soon as the data is available and 5) push the next timestep to the remaining open buffer. 6) Once a timestep is released on the reader, a receive buffer is available, and this 7) is advertised to the writer ranks in metadata.}
    \label{fig:rdma_preload}
\end{figure}

The LibFabric data plane takes a somewhat different approach to {\bf LearnedPreloadMode}. Figure \ref{fig:rdma_preload} illustrates this approach. 
Each reader rank maintains a \texttt{RequestLog}, which is a history of all remote read calls made by that reader prior to selections being locked and preload pulls beginning. This log is kept with a three-level index: first, the timestep of the request; second, the writer rank the request is made from; finally, the order in which the request was received. The request log for a given \texttt{\{timestep, rank\}} pair is kept as a contiguous buffer, expanding at a growth factor of two, as necessary. If no entries are made for a particular \texttt{\{timestep, rank\}} pair, no buffer is ever allocated for that pair. Only the most recent timestep's request log is made available to the writer when preload selections are sent; previous timestep's reads are not used for preloads.

On the writer side, each writer rank creates a \texttt{ReaderRoll} buffer, which is large enough to hold one set of buffer information (an \texttt{\{address, length, access\_key\}} tuple) for each reader rank. It makes this buffer available for remote RDMA writes, and advertises the buffer to the readers during the \texttt{initPerReader()} process. When the preload is enabled, the readers make their \texttt{RequestLog} available for RDMA reads, and then write buffer access information for each non-empty \texttt{RequestLog} into the \texttt{ReaderRoll} of the respective writer, offset by the reader's rank. This is done as part of the first \texttt{EndStep()} after enabling preload, and the readers block on the completion of this write. In this way, the writer is guaranteed that all the \texttt{ReaderRoll} information has been populated by the time it is handling the timestep release. The writer reads the collated \texttt{RequestLog} from the readers and blocks the timestep release until these reads complete.

Each reader ranks maintains two preload buffers, each large enough to contain the preload for a single timestep. This is, in effect, a preload queue of depth two. As the writer progressively pushes timesteps, it alternates between preload buffers. The goal is for the preload pushes to stay a timestep ahead of the reader. This allows the next step to be pushed simultaneous with reads completing for the current timestep, and prevents the writers from becoming blocked waiting for a timestep to release.

The availability of a buffer as a preload target is affected by two events: a push being sent to the buffer (which makes the buffer unavailable) and the reader-side release (which makes the buffer available again.) Since the push is initiated on the writer side, and the control plane notifies the writer-side data plane of release events, the writer is able to maintain its own view of buffer usage on the reader. This view is conservative, in that there is some latency from the time a reader makes a buffer available by releasing its timestep and when the writer is notified of this. Thus, the writer can safely act upon its view of the reader's preload buffers without risking corrupting data on the reader.

When read selections are initially posted by the reader, the writer will push two timesteps, if they are available to be pushed, and update its view of the reader preload buffers to indicate they are unavailable. Subsequently, whenever a reader releases a timestep, the writer will send the next timestep if it is available, and mark the buffer available otherwise. Whenever a timestep is provided by the user it will be pushed immediately if there is an available preload buffer.

\subsubsection{Waiting for Readers and Step Distribution Options}
Section~\ref{sec:control_plane} notes that a writer can have multiple readers registered, but is purposefully vague about details to avoid complicating the narrative.
For example, the most common situation is for a writer (producer) to expect a single reader (consumer) for its steps, and want to not lose any steps (or have a failure) if their startup is not perfectly synchronized.
Therefore with the default settings and using files to exchange contact information:
\begin{itemize}
    \item on the reader side, \Open will poll for a configurable period of time ({\bf OpenTimeoutSecs}, default 60) waiting for the writer to start and produce its contact file.
    \item on the writer side, \Open will block waiting for a configurable number of readers to join ({\bf RendezvousReaderCount}, default 1).
\end{itemize}
This allows for specialization to support different circumstances.  
When a long-running simulation has an output stream that allows periodic connection for progress checking, {\bf RendezvousReaderCount} can be set to zero so that \Open doesn't block and produced steps will be managed as described in Section~\ref{sec:queue_management}.

\begin{figure}[htb]
    \centering
    \includegraphics[width=\columnwidth]{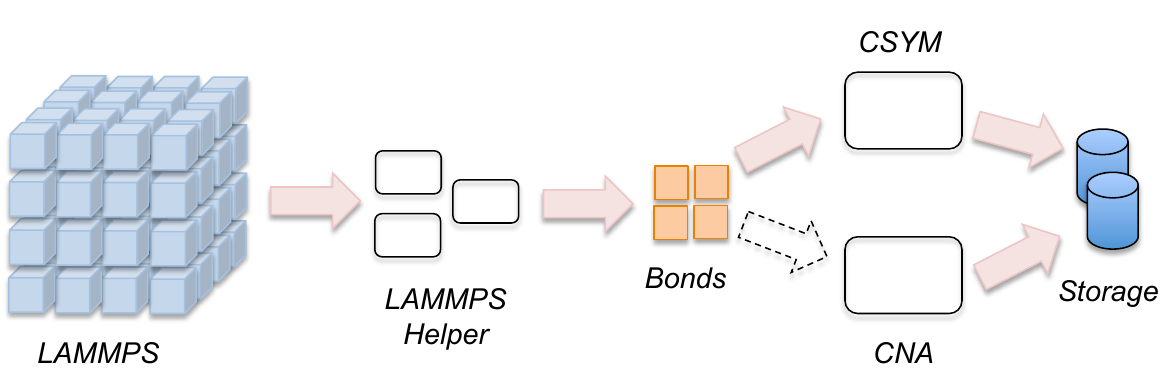}
    \caption{LAMMPS Analysis Pipeline.}
    \label{fig:split_workflow}
\end{figure}
Conversely, in a workflow like in Figure~\ref{fig:split_workflow}, {\bf RendezvousReaderCount} can be set appropriately for the number of consumers downstream from each workflow member.  In this sort of environment, it is relatively common that when multiple readers are present, each wants to see and select data from every step that the writer produces. 
However, this isn't true for every application, so SST has a {\bf StepDistributionMode} parameter to control how steps are distributed amongst registered writers.  
The parameter defaults to {\bf StepsAllToAll}, producing the "every reader gets every step" described above, but setting it to {\bf StepsRoundRobin} results in each step that the writer produces being sent to only a single reader in round-robin order.  In terms of the implementation details described in Section~\ref{sec:control_plane}, with {\bf StepsRoundRobin} the {\tt ProvideMetadata} message is sent to only the selected reader, not to any other registered readers, and the reference counts associated with writer data and metadata are set accordingly.  
The remainder of the reader/writer step management protocol remains unchanged.  

While {\bf StepsRoundRobin} is generally useful for load balancing between multiple consumers, it can be insufficiently flexible when the reader-side per-step processing time varies from step to step.  
In that circumstance, a reader that happens to receive a series of steps with high processing time might end up with several unprocessed steps in its metadata queue while other more lightly loaded readers are idle.
To address this situation, SST also has a {\bf StepsOnDemand} setting for the step distribution mode.  With this setting, {\tt ProvideMetadata} messages are not automatically sent to any reader in writer-side \EndStep.  
Instead, reader-side \BeginStep results in a {\tt RequestStep} message being sent to the writer.  
When this message arrives at the writer, if it has a step that has not previously been sent to any reader, it responds with a {\tt ProvideMetadata} message and reader-side \BeginStep proceeds as before.  
If the writer has no unprocessed steps available, the {\tt RequestStep} message is queued until the writer produces another step.  
In the event of multiple requests in the queue, new steps are assigned to readers in FIFO order, one event per request.  This protocol is a essentially a more general on-demand work-distribution method, matching each outgoing step to the first available consumer.  

\section{Performance Results}

Having described the goals, architecture and implementation of SST above, this section examines different aspects of SST's performance, comparing to file-based approaches, comparisons between different data planes and examining SST's behaviour on current Exascale platforms.

\subsection{WRF and the UCX data transport}
Our first study provides an examination of the ADIOS2 backend as used by WRFv4.5.0. The well-known benchmark provided by NCAR is a 6-hour simulation over the continental United States (CONUS) on a 1501x1201 grid at a 2.5km resolution. The experiments were conducted on an 8-node cluster with two 18-core Intel Xeon Gold 6240 CPUs, 384 GB DDR4 memory, and a Mellanox ConnectX-6 interconnect. The storage setup included a dedicated storage node with a BeeGFS file system and eight 10K RPM spinning hard disk drives connected to compute nodes using Mellanox ConnectX-5 NICs. Every computational node had an Intel DC P4510 1TB NVMe SSD disk. Advancing high-performance computing and data management relies heavily on the development of sustainable and efficient transport engines. 
The UCX data plane is a relatively recent addition to SST, proving another RDMA-capable data plane to expand SST's capabilities to other class of HPC network hardware.  Prior to the creation of the UCX data plane, the only data plane available on this hardware was the EVPath TCP-based data plane because libfabric was not supported on this hardware and the MPI data plane of SST did not exist yet.
\begin{figure}[htb]
    \centering
    \includegraphics[width=0.48\textwidth]{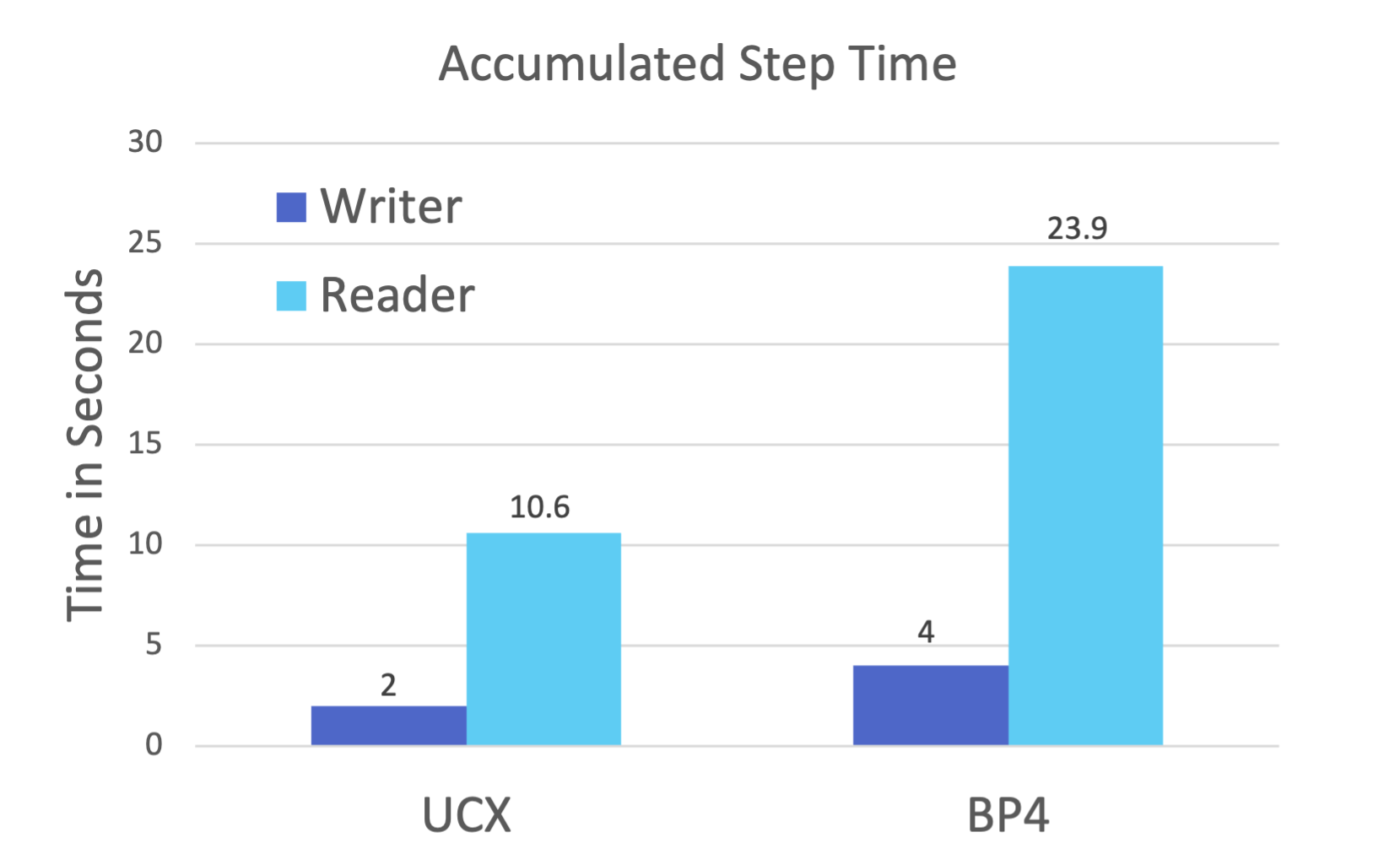}
    \caption{A WRF run with postprocessing is compared in terms of run time. The graph shows that the ADIOS streaming engine outperforms the file-based ADIOS2 engine BP4 significantly on both the Reader and Writer sides}
    \label{fig:ucx1}
\end{figure}
To evaluate the efficiency of UCX transmission, we increased the frequency of 2.6 GB WRF history file production of the benchmark to once every 30 minutes of simulated time, which was an appropriate time scale for data analysis. Figure~\ref{fig:ucx1}  compares the I/O cost of the benchmark run, as the simulation (writer) and analysis (reader) perceive it, between file-based (BP4) and SST/UCX staging. The results indicate that the UCX transport, serving as a low-level communication library, introduced a more efficient and flexible interface for managing data communication across various nodes. 

\begin{figure}[htb]
    \centering
    \includegraphics[width=0.48\textwidth]{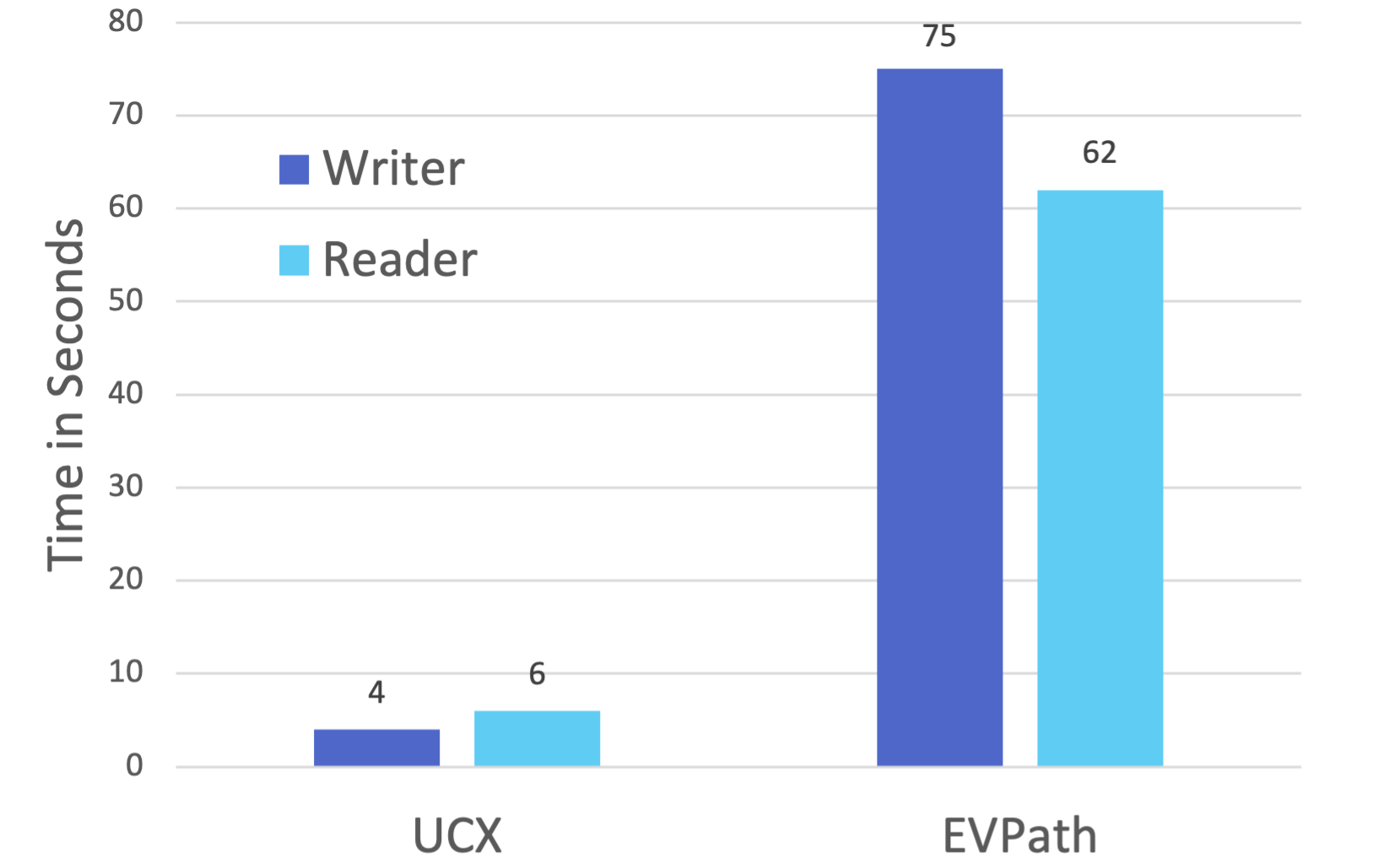}
    \caption{A WRF run with postprocessing was executed and compared in terms of runtime using 4 nodes. The ADIOS2 SST engine employing UCX data transport (left) significantly outperformed the default EVPATH data transport (right) by a factor of ten, with QueueLimit set to 1.}
    \label{fig:ucx2}
\end{figure}
The UCX transport also surpassed the EVPath data plane leading to swifter data transfer rates and a smoother user experience. This performance boost is evident in another test case in Figure~\ref{fig:ucx2}, which shows the UCX transport outperforming EVPATH in terms of both throughput and latency because of using RDMA for asynchronous data movement instead of a slow synchronized back-and-forth using TCP. The UCX transport can accommodate a significantly higher volume of data while transmitting data with considerably lower latency compared to EVPATH. 
Overall, the integration of the UCX transport into the SST engine has greatly enhanced its capabilities on this hardware, ensuring that the SST engine remains a valuable tool for researchers and scientists. The UCX transport has enabled ADIOS2 to demonstrate the significant benefits of merging cutting-edge open-source libraries like ADIOS2/UCX with classic HPC applications like WRF.



\subsection{WDMApp}
\label{sec:wdmapp}

\begin{figure}[htb]
    \centering
    \includegraphics[width=0.48\textwidth]{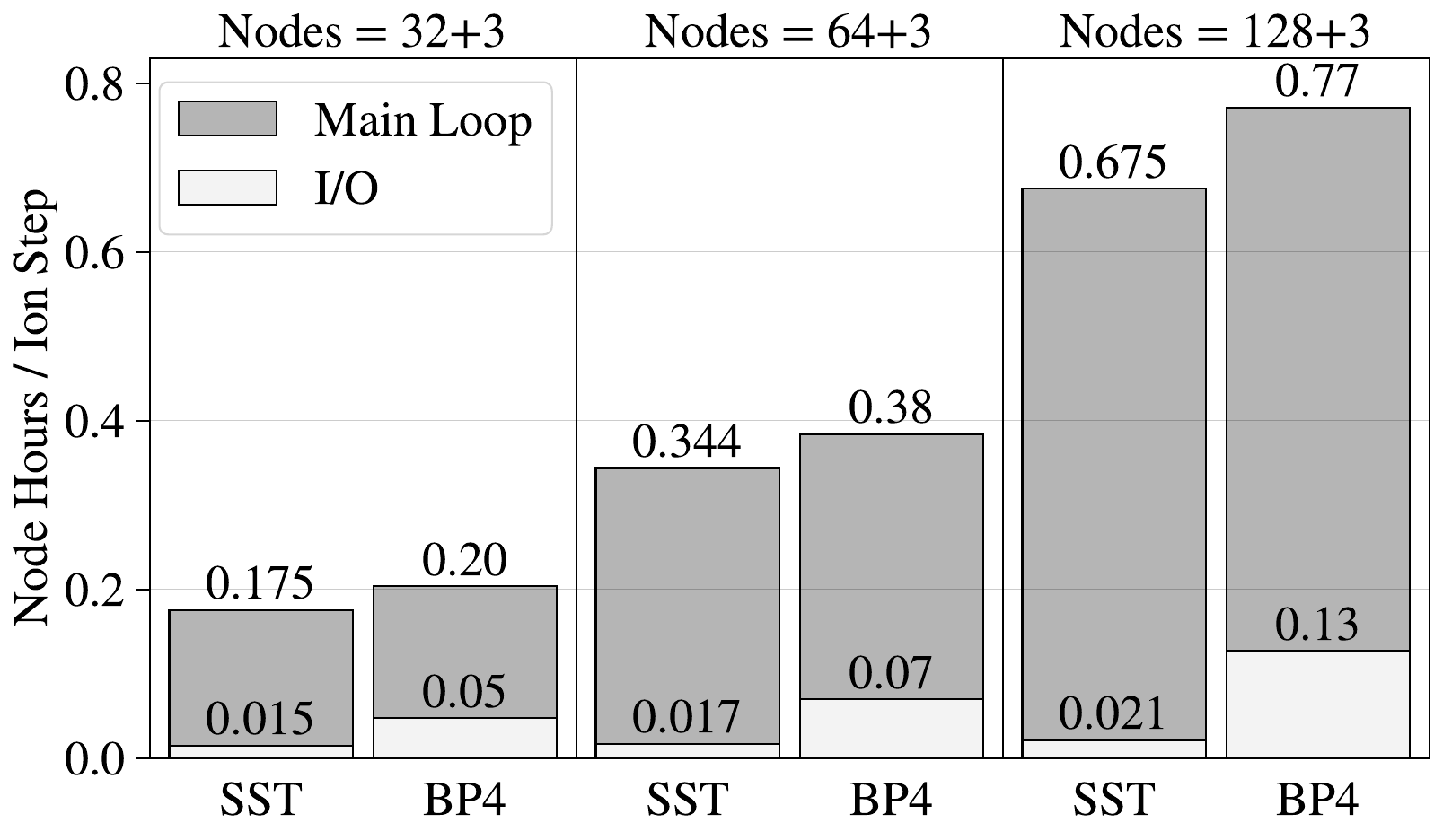}
    \caption{Results of inter-code data movement I/O overhead in XGC-GENE coupling on Crusher, compared to the main loop cost. SST was configured with the MPI method, and significantly reduced the overhead compared to coupling through files. Nodes counts at the top of the figure are XGC+GENE. This is a weak scaling run, where XGC runs for about the same wall-clock time when doubling its size and the number of particles.}
    \label{fig:wdmapp-overhead}
\end{figure}

Whole device modeling (WDM) is the effort of the plasma physics community to predict \textit{all} the relevant multi-scale physics of a fusion device in a consistent, robust fashion. Characteristic phenomena span several orders of magnitude on both spatial and temporal scales. Because certain models, representations, and approximations are more valid and/or efficient in certain regimes than others, various HPC applications have been specialized for simulations of a subset of the full physics; a uniform, first-principles-based simulation on all scales is not computationally feasible. 

WDMApp was a project under the Exascale Computing Project (ECP), which adopted a multi-application approach to WDM, sharing data between separate codes with ADIOS. Exploring spatial coupling in a fluid model paradigm~\cite{DominskiSpatialCoupling}, WDMApp ran XGC-GENE~\cite{GENE-XGC-Coupling} and XGC-GEM~\cite{GEM-XGC-Coupling} workflows. XGC includes the physics needed to resolve edge turbulence in complex geometries, while GENE and GEM are two alternatives which use different approaches appropriate for the core region of the device. The core-edge data exchange here constitutes \textit{strong coupling}, where field data on the mesh must be sent and received multiple times in each direction per (ion) simulation time step. Accordingly, it is important to achieve performant enough data movement between the codes such that this is not a bottleneck.

\autoref{fig:wdmapp-overhead} plots WDMApp coupling performance for XGC-GENE workflows on Crusher, Frontier's testbed during the ECP project. The data movement between the codes proceeded using either BP4 or MPI-configured SST, with SST achieving significant savings in time and scaling penalty. These results are updated from previous WDMApp performance measurements using SST in an earlier version of the code coupling algorithm, which used less data exchange~\cite{EFFIS}. The code coupling of fusion applications in this project ultimately did not pose a performance challenge for ADIOS. More interesting use of SST by this project is in situ analysis of XGC data, using 1 extra compute node to produce three different analyses for a large simulation on 1024 nodes, i.e., at marginal cost to the user. These plasma physics workflows are described in ~\cite{SuchytaCluster}.

\subsection{In-Transit Machine Learning of Plasma Simulations}

When using physics simulation data as an input for Machine Learning (ML), the high volume of data produced by state-of-the-art HPC simulation codes combined with the large amounts of data typically required for training neural networks result in a data challenge that is hard to overcome for conventional IO solutions, especially for workflows targeting full scale execution on modern HPC systems.
Since simulation and ML codes usually employ vastly different software stacks and the explorative nature of machine learning requires outstanding flexibility and adaptability, such data simulation pipelines are usually modeled as \emph{loosely-coupled} setups\cite{PoeschelSMC2021} in which both codes are developed and executed separately, feeding the ML code with data produced by the simulation code.
This observation amplifies the data challenge since simulation data needs to be transmitted into the ML code in a portable, flexible and scalable way.

In a study presented in Ref. \cite{sc2024},  a particle-in-cell (PIC) simulation of the Kelvin-Helmholtz instability (KHI) computed by the GPU-accelerated PIConGPU code \cite{Bussmann2013} is coupled with a PyTorch-based \cite{NEURIPS2019_9015} training code (TC) to train a data-driven model for reconstructing the phase space from the  particle data and the in-situ computed radiation data of the simulation.  
PIConGPU produces its output in terms of the openPMD standard \cite{huebl2015openpmd}, using the openPMD-api \cite{openPMDapi} that itself builds upon ADIOS2. 
openPMD is a data standard for \textbf{p}article-\textbf{m}esh-\textbf{d}ata built for F.A.I.R.\ scientific IO in HPC software that needs to scale. 

In the described loosely-coupled setup, the SST engine of ADIOS2 plays a major role for overcoming the IO bottleneck by avoiding the parallel file system, replacing file IO with streaming IO.
In particular, there is no lock-in to a specific pattern for data exchange, compared to coupling approaches based on e.\,g.\ local SSDs or shared memory. While local data exchange within a node remains preferred for scalability, this IO approach naturally extends towards patterns such as staging within a neighborhood of nodes (for scheduling reasons or for implicit load balancing via streaming) or a fan-in pattern (for data reduction purposes).

Deployment on OLCF Frontier can be considered via either the MPI data plane, based on MPI's inter-communicators between separate MPI worlds, or via the LibFabric data plane based on the CXI provider \cite{libfabric_github}, this way presenting the chance to compare performance numbers between different implementations. With the more low-level LibFabric, further performance-impacting requirements by the MPI data plane can be circumvented, such as the use of threaded MPI or required writer handshake interaction for remote read accesses. While a LibFabric data plane exists previously in ADIOS2, CXI requires adding support for (1)~the \verb|MR_ENDPOINT| memory mode, (2)~non-\verb|MR_VIRT_ADDR| memory mode, (3)~manual data progress mode and (4)~CXI-specific authentication logic.


In the context of a weakly-scaling loosely-coupled setup, an IO solution is required that can scale to the full scale of exascale systems such as Frontier. For verifying that SST can provide this, a synthetic benchmark is presented that reads particle data produced by PIConGPU's KHI scenario (5.86~GB per compute node and time step) into a no-op data sink that does nothing other than measuring the time for loading the data. PIConGPU is executed on all GPUs, i.\,e.\ using 8 MPI ranks per compute node, while the data sink is executed concurrently to the simulation with one MPI rank per compute node on the CPU, collecting the node's data. By computing the throughput based on (1)~this measured time and (2)~the total transferred data size, the resulting value -- the perceived throughput -- also includes the overhead for communication and synchronization and hence is a lower bound on the actual throughput.

\begin{figure}[htb]
\centering
\includegraphics[width=\linewidth]{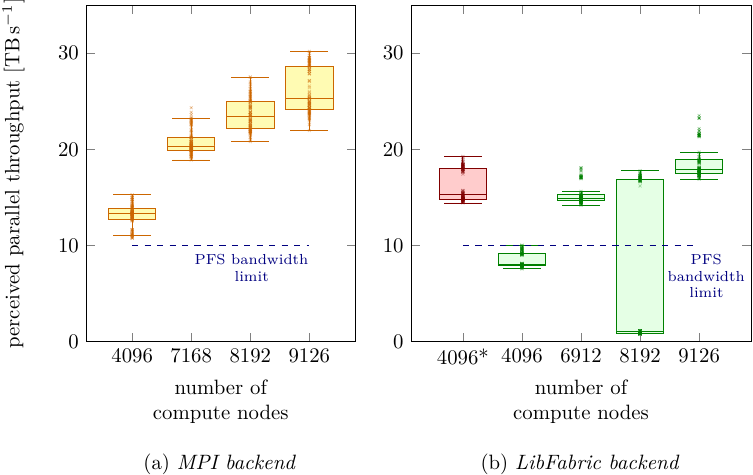}
\caption{\textit{Boxplots of the parallel throughput for streaming at full scale on Frontier using a synthetic benchmark built on PIConGPU KHI. (a)~Using the MPI data plane. (b)~Using the LibFabric data plane. (One single run, marked \(4096\ast\), used a configuration that achieved a higher throughput, but did not scale to the full system.)
The full-scale throughput reaches between 20 and 30~TB/s for MPI, and around 20~TB/s for LibFabric.}
}
\label{fig:synthetic-benchmark}
\end{figure}

The benchmarks presented in Fig.\ \ref{fig:synthetic-benchmark} 
show that streaming IO via ADIOS2 and SST scales well to OLCF Frontier, the current TOP1 HPC system, depicted as boxplots of all single measurements. The perceived parallel throughput of 20 to 30~TB/s at full scale exceeds the 10~TB/s theoretical peak bandwidth of the parallel Orion filesystem, a scaling limit circumvented by not using the filesystem. Even a comparison with the 35~TB/s aggregate write bandwidth of the SSDs installed locally in the compute nodes\cite{frontier_pfs_bandwidths} demonstrates that SST sets loosely-coupled applications up for IO at the exascale.

The results show solid throughput for the MPI data plane, while the LibFabric data plane allows, but also requires further finetuning: The highest per-node throughput of all benchmarks ($3.5\sim$~4.7GB/s) is reached in a single run of the LibFabric data plane at 4096 compute nodes (labeled $4096 *$ in Fig.~\ref{fig:synthetic-benchmark}). In here, the read operations per step were enqueued all at once, a strategy that results in a high performance, but that did not scale to the entire system due to overloading the network. To avoid this, throttling measures became necessary, in this study by submitting operations in batches of 10. This workaround brought the per-node throughput down to $1.9\sim2.6$~GB/s. The measured per-node throughput for the MPI data plane ranges between $2.6\sim3.7$~GB/s at 4096 compute nodes to $2.4\sim3.3$~GB/s at 9126 nodes. All single data transfers except one outlier of the LibFabric implementation complete within 1.2~-~3.2s


In order to tell the difference between perceived throughput and actual throughput, the same setup is executed again on 2 nodes (Frontier does not initialize its networking in single-node jobs) in a series of benchmarks that continuously doubles the simulation resolution. This leads to a benchmark series in which the overhead $t_\mathrm{overhead}$ is a constant time since not the number of datasets, but only their sizes increase, leading to an increasing loading time $t_\mathrm{load}$. The real throughput then depends on the measured time via: 
\[ t_\mathrm{total} = t_\mathrm{load} + t_\mathrm{overhead} = \tfrac{\mathrm{data}}{\mathrm{throughput}} + t_\mathrm{overhead}\]

\begin{figure}[htb]
    \centering
    \includegraphics[width=0.7\linewidth]{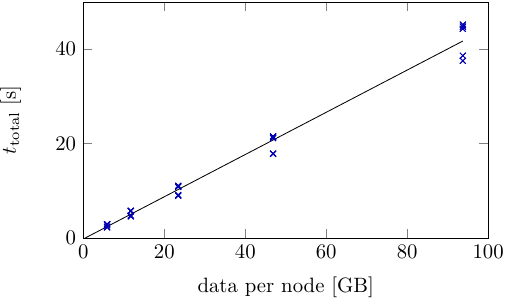}
    \caption{Scaling the simulation resolution in a 2-node setup. Linear regression yields a $y$-axis intersection at $t_\mathrm{overhead}=-0.17$~s and a coefficient of $\frac{1}{\mathrm{throughput}}=0.45$~s/GB per compute node, i.\,e.\ a throughput of 2.24~GB/s.}
    \label{fig:pipe-singe-node}
\end{figure}

Scaling the data to different sizes results in a linear equation $t_\mathrm{total}(\mathrm{data})$ whose unknown constants $\frac{1}{\mathrm{throughput}}$ and $t_\mathrm{overhead}$ can be found via linear regression of the timing results, shown in figure \ref{fig:pipe-singe-node}.

Linear regression shows that at this scale, the constant overhead is practically negligible and the perceived throughput is indistinguishable from the real throughput, measuring roughly 2.24~GB/s per compute node.

As promising approaches for increased per-node throughput, either  (1) parallel SST streams per node are likely to saturate the communication infrastructure even better, or alternatively (2) the upcoming linkX provider of LibFabric \cite{pritchard2023open} can help with memory hierarchy awareness. This meta provider links the CXI provider and a shared memory provider of LibFabric into one common provider, solving the lacking use of intra-node communication techniques by the CXI provider. Since the  LibFabric data plane of SST now supports the CXI provider as well as the shm provider (for shared memory), it is very likely that ADIOS2 will be able to leverage linkX.

\section{Conclusion}
In exascale computing environments, the ability to stream directly from large-scale data producers to consumers, thus avoiding potential bottlenecks in the filesystem, is of growing importance.  
In this paper we have examined the advantages and challenges of utilizing the proven file-oriented ADIOS I/O system, adding an ``engine'' to stream data from writers to readers, and thus enabling existing ADIOS applications to exploit these direct streaming capabilities with few if any code changes.
We discussed the overall design, internal structure and operation of the Sustainable Staging Transport engine and presented studies of its performance to highlight different capabilities.  
These studies include a demonstration of scaling on the OLCF Frontier Exascale machine where SST achieves up to 30~TB/s of parallel throughput, exceeding by a factor of three the 10~TB/s bandwidth limit of the Orion filesystem, which shows that direct streaming of data can have significant advantages over filesystem-based approaches for linking computation and analysis.  
SST's data planes are still under development, and while the results we've demonstrated on Frontier using the MPI dataplane are satisfying, we have reason to believe that direct use of the underliying network using a tuned libfabric transport will achieve even better throughput.
\section*{Acknowledgments}
This research used resources of the Oak Ridge Leadership Computing Facility, which is a DOE Office of Science User Facility supported under Contract DE-AC05-00OR22725. 

The core code development was supported by the Exascale Computing Project (17-SC-20-SC), a collaborative effort of the U.S. Department of Energy Office of Science and the National Nuclear Security Administration.

Application outreach efforts by the ADIOS team was supported by the U.S. Department of Energy, Office of Science, Office of Advanced Scientific Computing Research, Scientific Discovery through Advanced Computing (SciDAC) program, under the “RAPIDS Institute”.

Funded by the European Union. This work has received funding from the European High Performance Computing Joint Undertaking (JU) and Sweden, Finland, Germany, Greece, France, Slovenia, Spain, and Czech Republic under grant agreement No 101093261. 

This work was partly funded by the Center for Advanced Systems Understanding (CASUS) which is financed by Germany’s Federal Ministry of Education and Research (BMBF) and by the Saxon Ministry for Science, Culture and Tourism (SMWK) with tax funds on the basis of the budget approved by the Saxon
State Parliament.

\bibliographystyle{plain}
\bibliography{ref}

\end{document}